# CHEZ PL: A Hyper-Extensible AI-Integrated Zero-Trust CIAM-PAM Framework for Enterprise Security Modernization

Shivom Agarwal
*Quantum Dynamics SAS*
Marseille , France
shivom.aggarwal@qd-corp.com

Shourya Mehra
*Quantum Dynamics SAS*
Marseille , France
shourya97mehra@gmail.com

Safeer Sathar
*Quantum Dynamics SAS*
Marseille , France
safeervs7@gmail.com

*Abstract*— Customer Identity and Access Management (CIAM) systems play a pivotal role in securing enterprise infrastructures. However, the complexity of implementing these systems requires careful architectural planning to ensure positive Return on Investment (RoI) and avoid costly delays. The proliferation of Active Persistent cyber threats, coupled with advancements in AI, cloud computing, and geographically distributed customer populations, necessitates a paradigm shift towards adaptive and zero-trust security frameworks. This paper introduces the **C**ombined **H**yper-Extensible **E**xtremely-Secured **Z**ero-Trust (**CHEZ**) CIAM-PAM architecture, designed specifically for large-scale enterprises. The CHEZ PL CIAM-PAM framework addresses critical security gaps by integrating federated identity management (private and public identities), password-less authentication, adaptive multi-factor authentication (MFA), microservice-based PEP (Policy Entitlement Point), multi-layer RBAC (Role Based Access Control) and multi-level trust systems. This future-proof design also includes end-to-end data encryption, and seamless integration with state-of-the-art AI-based threat detection systems, while ensuring compliance with stringent regulatory standards.

This research paper outlines the architectural components of the CHEZ PL CIAM-PAM model, including its modular design, dynamic policy enforcement, and real-time monitoring capabilities. The proposed framework effectively minimizes technical debt during IAM migration processes, enabling smooth transitions from legacy systems while maintaining business continuity. Additionally, this paper evaluates the suitability of the CHEZ PL CIAM-PAM model for global organizations operating in distributed environments with diverse regulatory requirements and highlights strategies to mitigate integration challenges. The paper also explores potential enhancements, including the deployment of AI/GenAI tools for advanced risk analysis, behavior-based anomaly detection, and predictive analytics. By leveraging AI within a zero-trust architecture, the CHEZ PL framework ensures scalability, adaptability, and proactive security monitoring. Finally, the research concludes with insights into limitations, practical implications for enterprise adoption, areas for future development, and considerations for managerial decision-making.

*Keywords*— *CIAM architecture, CIAM-PAM, Zero-trust CIAM, password-less, AI-enhanced security, AI CIAM.*

## I. Introduction

The regular emergence of novel cybersecurity threats necessitates that organizations continuously adapt their security strategies to have an effective & efficient enterprise cybersecurity system. Customer Identity and Access Management (CIAM) and Privileged Access Management (PAM) systems have emerged as critical components for securing the enterprise systems. Typically, legacy IAM systems facilitated the definition and implementation of access control policies, determining authorized users, specific resource access, temporal restrictions, and contextual conditions. However, these traditional IAM solutions often fail to address modern requirements such as distributed and open Customer identities, scalability, password-less authentication, AI-driven monitoring, and zero-trust security models required for globally distributed systems.

The proposed framework leverages adaptive authentication, identity federation, AI-powered continuous session risk analysis, and integrated privileged access management system. Departing from traditional perimeter-based security policies, this architecture adopts a Zero Trust framework, enabling granular access control, real-time session monitoring, and adaptive policy enforcement. While traditional systems are constrained by static access control policies and a hierarchical access framework, this architecture enhances resilience through decentralized authorization and dynamically adapts access privileges contingent on contextual risk factors.

Key innovations of this framework include enhanced encryption standards, pseudonymized data sharing, federated identity management, and advanced behavioral analytics. These innovations support distributed network environments and multi-platform integration while adhering to compliance standards such as GDPR, HIPAA, and SOC 2. Additionally, the CHEZ PL CIAM-PAM model provides seamless interoperability with legacy systems due to underlying microservices design, facilitating enterprises to upgrade CIAM systems without causing major disruptions to existing workflows.

Furthermore, CHEZ PL integrates AI-driven anomaly detection for privileged access, session monitoring, and proactive risk assessments. It also employs real-time auditing tools that simplify compliance reporting and reduce overheads associated with manual audits. This research also investigates the efficacy of the proposed architecture in mitigating challenges related to scalability, interoperability across diverse platforms, and adherence to regulatory compliance standards, while also establishing a future-proof framework readily adaptable to emerging Artificial Intelligence (AI) and Generative AI (GenAI) technologies.

This paper concludes with a discussion on future extensions of this research paradigm, limitations of adopting CHEZ PL CIAM-PAM architecture and implications for Cybersecurity managers in large multi-national organizations.

## II. Literature Review

The paradigm of IAM systems has been changing over the last decade with more requirements of integrating external Customer identities, provide privileged access and demands of enhancing the monitoring & reporting. Historically, IAM architectures tend to be either role-based access control (RBAC) model, attribute-based access control model, mandatory access control model or discretionary access control model (Nahar & Gill 2022). But latent research (Yang et al 2014, Mohammed et al 2018, Nahar & Gill 2022) found that RBAC based architectures are most efficient for large scale organizations. Furthermore, these traditional IAM systems primarily relied on access control processes based on static roles and organizational hierarchies. The proliferation of cloud computing and geographically distributed customer populations, multiple open-source identities, active persistent cyber threats, etc. revealed limitations in existing architectures, highlighting the need for more dynamic, scalable, and adaptive solutions. Zero Trust architecture design, initially popularized by Forrester Research, has become a cornerstone of latest Identity and Access Management (IAM) systems (Turner et al 2021).

Besides, Zero-trust implications, IAM policies need to adapt to the emerging trend of multi-identity dynamics & data privacy compliance rules for external users, including – open identities, state-sponsored identities, social media identities, other trusted third-party identities (Peterson et al 2008, Roy 2023, Glöckler et al 2024). To address this challenge, federated identity models are increasingly utilized in contemporary deployments of CIAM to optimize authentication workflows across hybrid and multi-cloud environments (Malik et al 2015, Pöhn & Hommel 2020, Kiourtis et al 2023). The resulting federated identity systems centralize identity verification, alleviating the need for users to manage disparate credentials, thus enhancing both usability and security. Standards such as SAML (Hughes & Maler 2005), OAuth 2.0 (Fett et al 2016), and OpenID Connect (Mainka et al 2017) also significantly facilitated the interoperability between CIAM and PAM solutions across heterogeneous platforms (Walker 2019).

Other scholars have proposed to enhance existing IAM systems by incorporating password-less authentication (Alqubaisi et al 2020), AI-driven identity governance (Azhar 2016, Hawa 2024), risk-based access control models (Atlam et al 2020). Furthermore, the advancements in AI are also impacting our understanding of IAM systems, especially for automation and control. Many studies have postulated different methodologies to incorporate IA into IAM architectures to - authentication, authorization, and auditing (Aboukadri et al 2024, Vegas & Lamas 2024, Ahmadian et al 2014).

Incorporating all these different paradigms into one holistic architecture is crucial research gap and requires deep knowledge of cybersecurity architectures as well as AI systems design. This paper takes this challenge to bring all these state-of the-art technologies into CIAM architecture. This CHEZ PL CIAM-PAM architecture advances beyond this prior work by implementing a synergistic integration of Zero Trust principles and AI-driven threat detection, specifically designed to overcome scalability and compliance obstacles prevalent in large-scale enterprise environments It incorporates pseudonymized data models and compliance mechanisms for GDPR and HIPAA, ensuring regulatory alignment. Furthermore, it extends prior work by addressing integration challenges with legacy systems, providing backward compatibility, and enabling gradual migration strategies, in turn reducing overall project costs.

This paper extends prior work by offering a comprehensive, modular framework designed specifically for large-scale enterprises operating in geographically distributed networks. This architecture emphasizes interoperability with legacy systems while integrating adaptive authentication, AI-driven analytics, and policy enforcement mechanisms to effectively mitigate emerging cybersecurity threats and minimize technical debt. It highlights the practical application of federated identities and pseudonymized data models, ensuring compliance with regulatory standards across multiple jurisdictions. Finally, the framework leverages AI-driven monitoring for continuous improvements, adaptive threat response, and risk mitigation strategies.

## III. Proposed Methodology

The proposed CHEZ PL CIAM-PAM architecture is structured to address the limitations of traditional CIAM-PAM frameworks while ensuring scalability, security, and compliance. An architecture HLD design methodology for complex IT systems (Platunov et al 2014, Diaz-Pace & Bianchi 2019, Li et al 2024)

The methodology is divided into five primary phases:
- **Requirements Gathering and Analysis:** This phase identifies enterprise-specific security policies, needs & technical debt from legacy systems, business & compliance requirements, and existing gaps in CIAM-PAM systems. This involves stakeholder consultations, conducting risk assessments, and verifying data flow analysis to ensure the architecture aligns with business & cybersecurity objectives.
- **Framework Design and Architecture Planning:** A modular design approach is employed using microservices & incorporating multi-level authentication system, federated identity management, role-based access control, and AI-driven anomaly detection mechanisms. This phase also defines envisioned data encryption protocols, attributes pseudonymization techniques, and codifies integration points with existing systems.
- **Implementation and Integration:** The implementation phase involves deploying the CIAM-PAM framework, configuring access controls, integrating AI modules for monitoring, and enabling adaptive MFA mechanisms. Automated workflows for policy enforcement, identity federation, and data encryption are established.
- Testing and Validation: Comprehensive testing scenarios are executed, including penetration testing, vulnerability assessments, and stress testing. Performance metrics such as latency, throughput, and scalability are validated to ensure the architecture meets enterprise-grade requirements.
- Deployment and Optimization: The final phase involves rolling out the solution, conducting user training, and optimizing configurations based on feedback. Continuous monitoring tools and AI analytics are used to refine access controls, identify anomalies, and enhance performance.

This methodology ensures seamless migration from legacy systems, reduces operational risks, and enables dynamic scalability while maintaining compliance with regulatory standards. This research paper's scope covers only first 3 steps of requirements gathering, framework design & architecture planning and implementation & integration. The last two steps are left for a specific organization to deploy CHEZ PL CIAM-PAM architecture, test the integrated system, validate with power users & business managers, and enforce a continuous improvement process for regular optimization.

## IV. MODEL & ARCHITECHTURE

The Combined Hyper-Extensible Extremely-Secured Zero-Trust (CHEZ) CIAM-PAM architecture proposed in this paper is designed to address the evolving cybersecurity requirements of large-scale enterprises. It provides a multi-layered security approach that integrates Identity and Access Management (IAM) and Privileged Access Management (PAM) into a unified framework, enabling dynamic policy enforcement, zero-trust principles, and AI-powered threat detection. This section provides a detailed breakdown of the CHEZ CIAM-PAM architecture, focusing on its core components, design principles, and operational workflows.

Access Manager. FIM integration requires an Access Manager ensuring the operational readiness of required components, such as Web Logic Server and associated Identity Management (IdM) elements; registering a HTTP Server as a partner with Access Manager for resource protection; configuring Identity Federation as both a service provider (SP) and/or an identity provider (IdP) with Access Manager; and configuring Access Manager to delegate authentication to, or authenticate on behalf of FIM. Prior to undertaking these integration tasks, several key components must be installed, including application deployment server, HTTP server, Access Manager, Identity Federation tool, and a web server plug-in for Access manager (required in authentication mode).

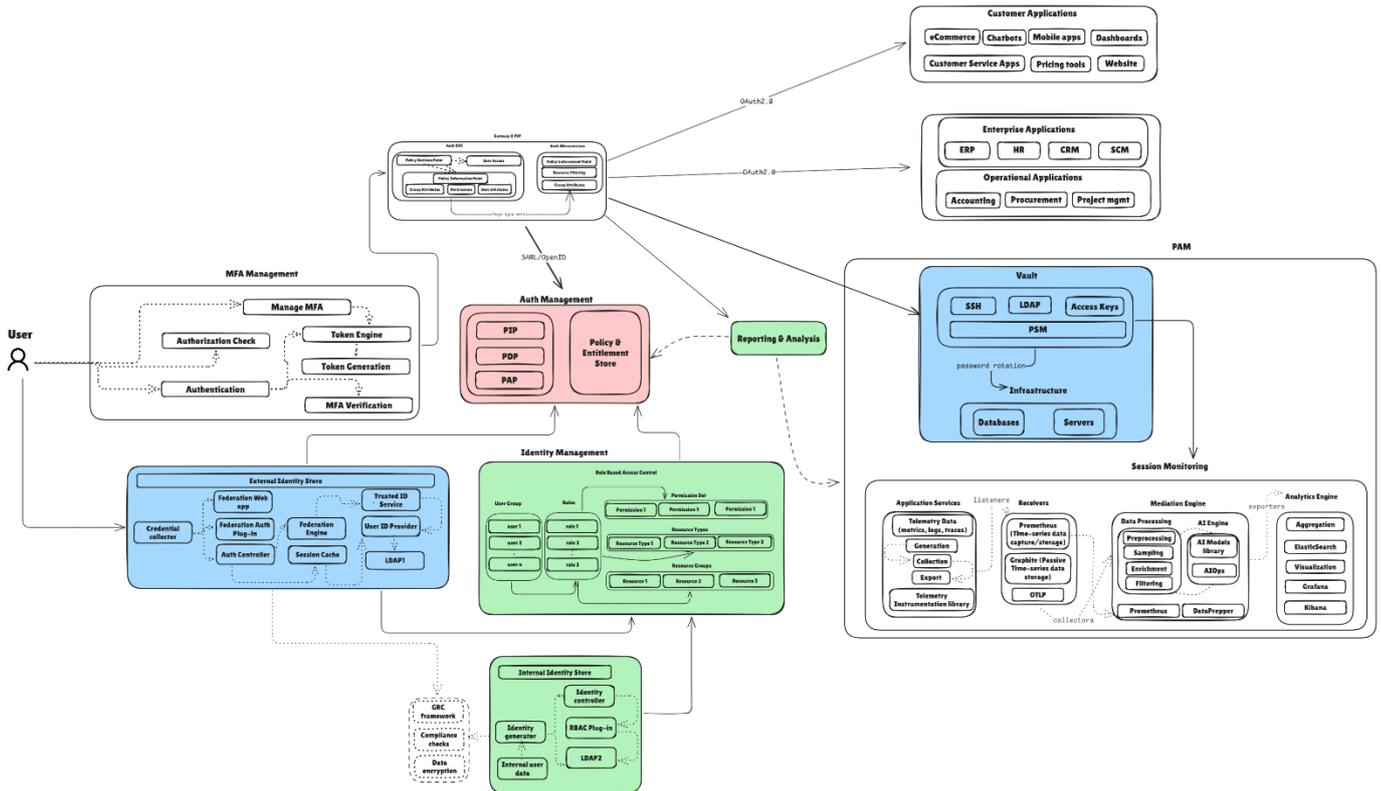

### A. Core Components of the CHEZ PL Architecture

#### 1. Identity Management Layer

**Federated Identity Management:** Supports multiple identity providers, enabling seamless integration of private and public identities. It allows organizations to centralize authentication while enabling cross-domain identity federation, ensuring secure interoperability. CHEZ PL CIAM-PAM architecture enables the integration of disparate Identity Providers (IdPs) utilizing SAML 2.0 and OpenID Connect (OIDC) protocols, allowing for the use of federated user attributes in access control decisions.

FIM is core component of the architecture that can support multiple federation protocols including SAML 1.x and SAML 2.0. Key capabilities include cross-protocol single sign-on (SSO), native integration with Access Manager and interoperability with any LDAP directory supported by the

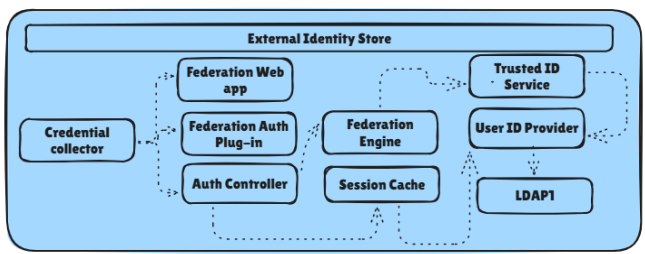

**Role-Based Access Control (RBAC):** Utilizes hierarchical roles and permission sets to define user access levels. The RBAC model ensures minimal privilege principles and enforces granular access policies.

In CHEZ PL CIAM-PAM architecture, RBAC has a defined component - The User Service, responsible for managing user-related functionalities as well as verifying user identities and granting access to protected resources. It serves as a central hub for user registration, and profile management, ensuring seamless interaction between users and other services within the system.

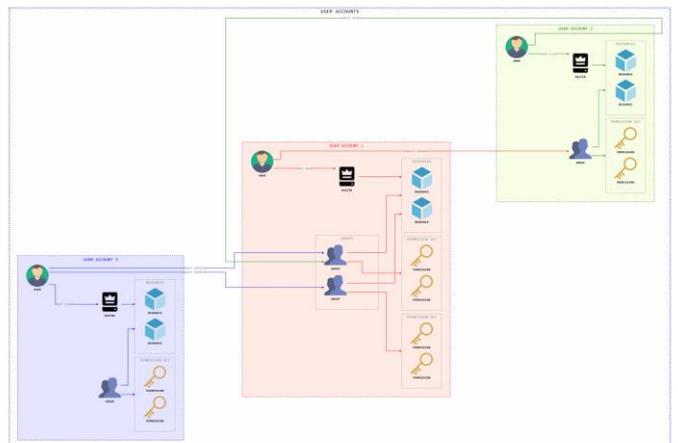

The architecture is centered around a master entity, serving as the primary entry point for the system. Users, groups, permissions, and resources are foundational components intricately linked to this master entity. Within each user account, a root user is granted the highest level of access and control. Groups, resources, and permission sets are associated with the master entity, defining their relationship with the account. Groups can also have specified permissions and resources, thus enabling fine-grained access control and resource management. Furthermore, users from one account may be members of groups in other accounts, facilitating collaboration and cross-account access management.

CHEZ PL CIAM-PAM architecture's design leverages several key patterns to enhance modularity and maintainability. The Singleton Pattern has been employed in the design through a "SingletonSessionManager" class to manage user sessions efficiently across the system. The Strategy Pattern provides further flexible authentication options for a given user by defining "PasswordHashingStrategy" and "EmailVerification Strategy" interfaces with concrete implementations such as bcrypt hashing and token-based verification. Moreover, the Observer Pattern enables event-driven behavior via an "UserEventListener" interface, allowing classes to respond to events like registration completion or password resets. The Decorator Pattern facilitates the dynamic extension of user profiles with features like social media links and themes, while also enabling the addition of functionality like logging and caching. Access control is enforced using the Proxy Pattern, implemented via an "AccessControlProxy" class that checks permissions before granting access to sensitive resources. Finally, the Builder Pattern provides a structured approach for constructing user objects, utilizing a "UserBuilder" class to manage attributes like name, email, and password.

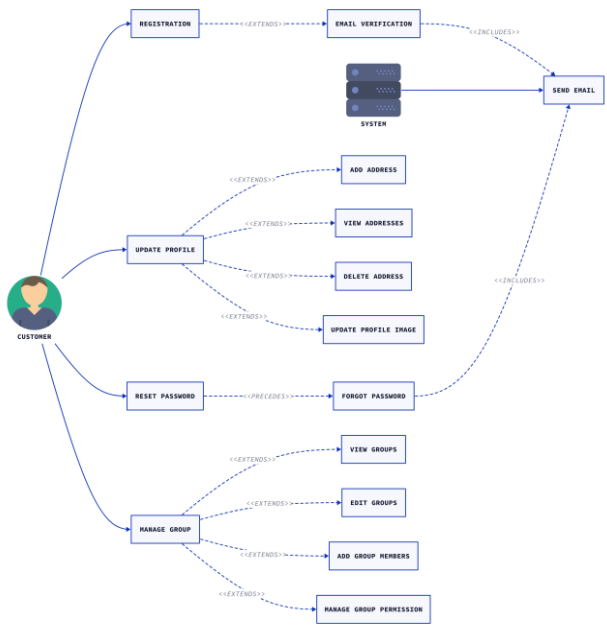

CHEZ PL's Identity management system architecture is characterized by a Customer actor, representing the user, and a System actor, representing the main application. The architecture supports several categories of use cases, including: account registration and verification (with registration, email verification, and send email functionalities), password recovery (comprising forgot password, reset password, and send email functionalities), profile management (including profile update, address management and image updates), and group management (with group view, edit, member addition, and permission management). Note that the delete address functionality has been disabled. In email verification and password reset processes, the *Send Email* action is invoked to deliver necessary communication.

*Identity Management LLD* - *A LLD (Low-level Design) implementation example for such a user flow process where a robust Captcha is also deployed.*

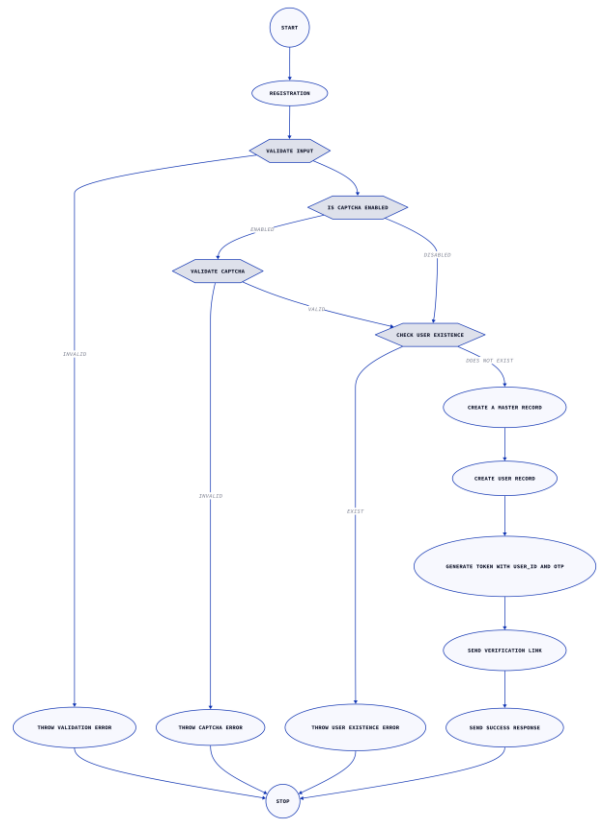

***Validation of Input (Condition)***: *The system rigorously validates the user-provided input, ensuring the integrity and accuracy of critical information including:*

- **Name**: Checks the name input for completeness, valid characters, and adherence to specified length limits.
- **Password**: Validates the password to meet predefined criteria such as length, complexity, and inclusion of special characters for enhanced security.
- **Date of Birth (DOB)**: Verifies the date of birth format to adhere to the specified pattern (e.g., dd/mm/yyyy) and validates against acceptable age ranges.
- **Email**: Ensures the email address follows the standard email format (e.g., user@example.com) and validates against common email validation rules.
- **Phone Number**: Validates the phone number format to conform to standard conventions and ensures it contains only numeric characters with optional country code.
- **Throw Validation Error (Process)**: If the input is invalid, an error is thrown, and the registration process stops.
- **Check for CAPTCHA Enablement (Condition)**: Determines whether CAPTCHA verification is enabled or not from the .env variable. If CAPTCHA is enabled, the system proceeds to "Validate CAPTCHA"; otherwise, it directly checks for user existence.
- **CAPTCHA Validation (Condition)**: Validates the CAPTCHA input. If the CAPTCHA is invalid, the process moves to "Throw CAPTCHA Error".
- **Throw CAPTCHA Error (Process)**: If the CAPTCHA is invalid, an error is thrown, and the registration process stops.
- **Check User Existence (Condition)**: Checks if the user already exists in the system. If the user exists with the

email or phone number, the process moves to "Throw User Existence Error"; otherwise, it proceeds to create a master record.
- **Throw User Existence Error (Process)**: If the user already exists, an error is thrown, and the registration process stops.
- **Create a Master Record (Process)**: Creates a master record if the user doesn't exist, preparing for user creation.
- **Create User Record (Process)**: Generates a new user record in the user table, associating it with the master record, with default role as "USER" and active status. Simultaneously, user-specific details, such as name, email, phone, password, and date of birth, are stored in the user_details table with initial verification flags set to false and null values for optional fields like profile image and OTP. password should be hashed using the bcrypt algorithm.
- **Generate Token with User ID and OTP (Process)**: Creates a token containing the user ID and one-time password (OTP) for verification purposes. Here we need to add one additional field to the token to identify the token type. Eg payload { userId: "user_123", otp: 123, type: VERIFY_EMAIL }. Token should be generated on Auth SVC.
- **Send Verification Link (Process)**: Sends a verification link containing the token to the user's email for account verification. Here, we can send the link, email and template id to the mail service.
- **Send Success Response (Process)**: Upon successful registration and verification, sends a success response indicating completion of the registration process.
- **Stop**: Marks the end of the registration process.

generated on Auth SVC.
- **Send Verification Link (Process)**: Sends a verification link containing the token to the user's email for password reset.
- **Send Success Response (Process)**: Upon successful sending of the verification link, sends a success response to the user.
- **Stop**: Marks the end of the forgot password process.

*Reset Password*
- **User Opens the Link (Process)**: Starts when the user opens the password reset link received via email.
- **Validate Token and Payload (Condition)**: Checks if the token and payload sent by the user are valid. Validate the token and payload on Auth SVC.
- **Throw Validation Error (Process)**: If the token or payload is invalid, an error is thrown, and the process stops.
- **Update the New Password in the Database (Process)**: If the token and payload are valid, updates the user's password in the database.And update the OTP in the database as NULL.
- **Send Success Response (Process)**: Upon successful password reset, sends a success response to the user.
- **Stop**: Marks the end of the password reset process.

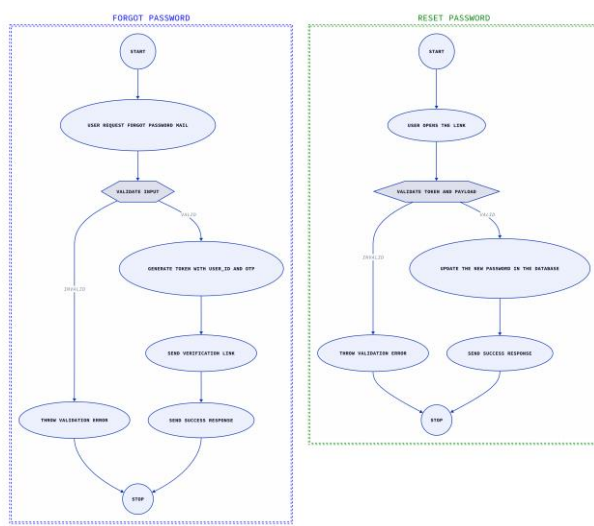

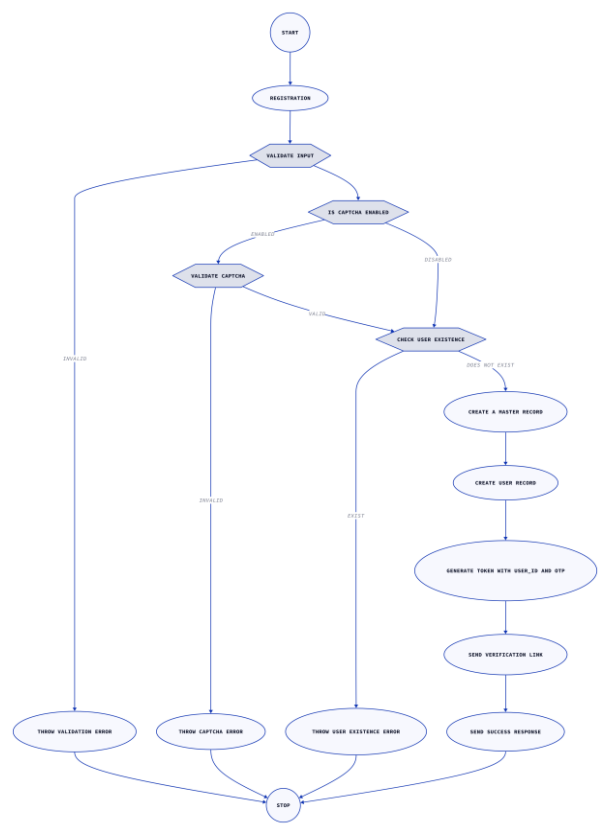

*Forgot Password*
- **User Request Forgot Password Mail (Process)**: Initiates the process when a user requests to reset their password.
- **Validate Input (Condition)**: Checks if the provided email and date of birth (DOB) are in the correct format and adhere to validation rules.
- **Throw Validation Error (Process)**: If the input is invalid, an error is thrown, and the process stops.
- **Generate Token with User ID and OTP (Process)**: Creates a token containing the user ID and one-time password (OTP) for verification purposes. Here we need to add one additional field to the token to identify the token type. Eg payload { userId: "user_123", otp: 123, type: FORGOT_PASSWORD }. Token should be

**Validation of Input (Condition)**: The system rigorously validates the user-provided input, ensuring the integrity and accuracy of critical information including:
- **Name**: Checks the name input for completeness, valid characters, and adherence to specified length limits.
- **Password**: Validates the password to meet predefined criteria such as length, complexity, and inclusion of special characters for enhanced security.
- **Date of Birth (DOB)**: Verifies the date of birth format to adhere to the specified pattern (e.g., dd/mm/yyyy) and validates against acceptable age ranges.
- **Email**: Ensures the email address follows the standard email format (e.g., *user@example.com*) and validates against common email validation rules.

- **Phone Number**: Validates the phone number format to conform to standard conventions and ensures it contains only numeric characters with optional country code.
- **Throw Validation Error (Process)**: If the input is invalid, an error is thrown, and the registration process stops.
- **Check for CAPTCHA Enablement (Condition)**: Determines whether CAPTCHA verification is enabled or not from the .env variable. If CAPTCHA is enabled, the system proceeds to "Validate CAPTCHA"; otherwise, it directly checks for user existence.
- **CAPTCHA Validation (Condition)**: Validates the CAPTCHA input. If the CAPTCHA is invalid, the process moves to "Throw CAPTCHA Error".
- **Throw CAPTCHA Error (Process)**: If the CAPTCHA is invalid, an error is thrown, and the registration process stops.
- **Check User Existence (Condition)**: Checks if the user already exists in the system. If the user exists with the email or phone number, the process moves to "Throw User Existence Error"; otherwise, it proceeds to create a master record.
- **Throw User Existence Error (Process)**: If the user already exists, an error is thrown, and the registration process stops.
- **Create a Master Record (Process)**: Creates a master record if the user doesn't exist, preparing for user creation.
- **Create User Record (Process)**: Generates a new user record in the user table, associating it with the master record, with default role as "USER" and active status. Simultaneously, user-specific details, such as name, email, phone, password, and date of birth, are stored in the user_details table with initial verification flags set to false and null values for optional fields like profile image and OTP. password should be hashed using the bcrypt algorithm.
- **Generate Token with User ID and OTP (Process)**: Creates a token containing the user ID and one-time password (OTP) for verification purposes. Here we need to add one additional field to the token to identify the token type. Eg payload { userId: "user_123", otp: 123, type: VERIFY_EMAIL }. Token should be generated on Auth SVC.
- **Send Verification Link (Process)**: Sends a verification link containing the token to the user's email for account verification. Here, we can send the link, email and template id to the mail service.
- **Send Success Response (Process)**: Upon successful registration and verification, sends a success response indicating completion of the registration process.
- **Stop**: Marks the end of the registration process.

*Other User Verification & Validation*
- **User Requests Email Verification (Process)**: The process initiates when a user requests to verify their email address.
- **Validate Email (Condition)**: Checks if the provided email address is in the correct format and adheres to validation rules.
- **Throw Validation Error (Process)**: If the email address is invalid, an error is thrown, and the process stops.
- **Generate Token with User ID and OTP (Process)**: Creates a token containing the user ID and one-time password (OTP) for verification purposes. Here we need to add one additional field to the token to identify the token type. Eg payload { userId: "user_123", otp: 123, type: VERIFY_EMAIL }. Token should be generated on Auth SVC.
- **Send Verification Link (Process)**: Sends a verification link containing the token to the user's email address for email verification.
- **Send Success Response (Process)**: Upon successful sending of the verification link, sends a success response to the user.
- **User Opens Verification Link (Process)**: The process begins when the user opens the email verification link received in their email inbox.
- **Validate Token and Payload (Condition)**: Checks if the token and payload sent by the user are valid. Validate the token and payload on Auth SVC.
- **Throw Validation Error (Process)**: If the token or payload is invalid, an error is thrown, and the process stops.
- **Verify Email and Update the Database (Process)**: If the token and payload are valid, verifies the user's email address and updates the database accordingly. And update the OTP in the database as NULL.
- **Send Success Response (Process)**: Upon successful email verification, sends a success response to the user.

**MFA Management**

Multi-factor authentication (MFA) bolsters security by mandating multiple authentication factors throughout the user lifecycle. The multi-factor authentication (MFA) process begins with *Authentication*, where the system verifies the customer's identity using credentials such as username and password. Successful authentication leads to *Token Generation* for subsequent resource access. Following authentication, an *Authorization Check* is performed to ensure the customer has permission to access the requested resource. If authorized, the process proceeds to *Multi-Factor Authentication (MFA)*, which adds an extra layer of security by requiring a secondary factor, such as a one-time code. Successful completion of the MFA prompts the generation of a second *Token Generation*, granting the customer access.

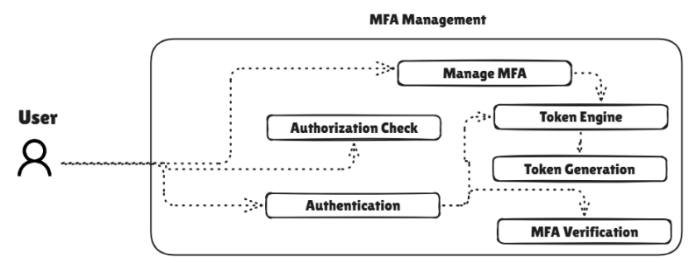

The MFA process can be broken down into key phases: a *Registration Phase* where users associate their accounts with multiple unique identifiers, such as mobile devices or authenticator applications. An *Authentication Phase* follows, which requires users to provide the primary credentials (password) followed by a secondary authentication response generated by the linked device. This might involve receiving a one-time code or a push notification, which, upon verification from the user in the *Reaction Phase*, grants access to the system. MFA may vary, for example, with two-factor authentication (2FA) by using only two factors, a third-party authenticator application for managing the secondary authentication, the use of biometric authentication, and adaptive authentication based on the device being used.

### MFA Implementation LLD

The system's login process begins with an Initiation, followed by Input Validation, where user-provided email (or phone number) and password are checked for accuracy. If the input is invalid, a Validation Error halts the process. Next, the system checks if CAPTCHA is enabled; if so, CAPTCHA Validation is performed, and failure generates a CAPTCHA Error. Following this, Credential Validation ensures the user's credentials are valid. Invalid credentials result in an Invalid Credentials Error. The system then determines whether MFA is Enabled for the user. If enabled, the system verifies the MFA Validation Type (email or Google Authenticator) and generates a one-time password (OTP) through Generate OTP, which is delivered to the user through Send Verification OTP. A JWT token is generated for the MFA request using Generated JWT Auth Token for MFA Request. A JWT token and a refresh token are generated through the Generated JWT Auth Token and Refresh Token once this is completed. Otherwise, if MFA is not enabled, a JWT token and a refresh token are directly generated. A Success Response concludes the login.

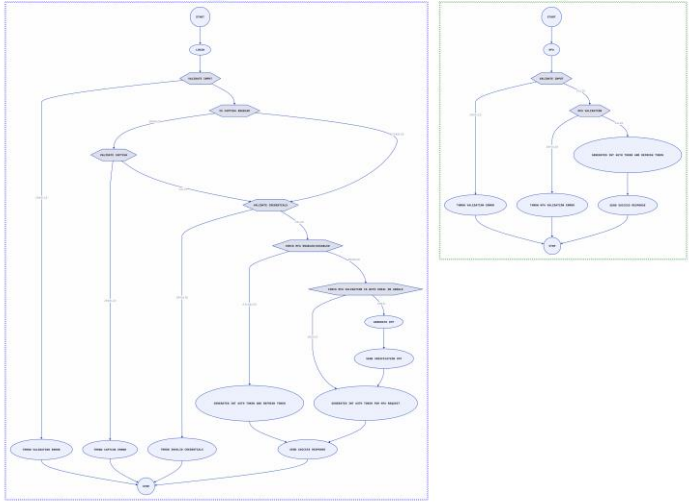

The Multi-Factor Authentication process starts with an Initiation, followed by Input Validation for the MFA OTP. An invalid OTP results in a Validation Error. The system then performs MFA Validation, throwing a MFA Validation Error for an incorrect OTP. If the validation is successful, a JWT token and a refresh token is generated through Generated JWT Auth Token and Refresh Token, followed by a Success Response to indicate the process is complete.

### Enable/Disable MFA Request
- **Validate Input**: Checks the validity of the input parameters provided by the user.
  - type
- **Throw Validation Error**: If the input is invalid, an error is thrown, and the process stops.
- **Type Check**: Determines whether the user requested to enable MFA via email or Google authenticator.
- **Generate OTP and Send Verification OTP Email**: If the user opts for email-based MFA, generates a one-time password (OTP) and sends it via email for verification. And it also saves the OTP in mfa table.
- **Generate Google Secret and QR Code Image**: If the user opts for Google authenticator-based MFA, generates a secret key and a QR code image for scanning. And it also saves the secret in mfa table. Generate OTP/ Google secret on AUTH SVC.
- **Send Success Response**: Upon successful generation and delivery of the verification mechanism, sends a success response to the user.

### Enable/Disable MFA
- **Validate Input**: Ensures the validity of the input parameters provided by the user.
  - OTP

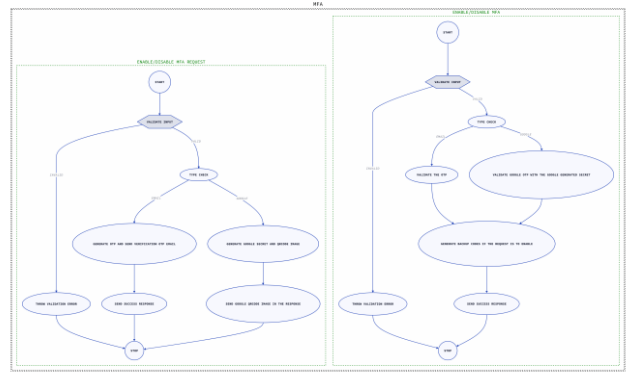

- **Throw Validation Error**: If the input is invalid, an error is thrown, and the process stops.
- **Type Check**: Identifies whether the user chose email-based or Google authenticator-based MFA.
- **Validate the OTP**: Fetch the mfa row by user id, If the user chooses email-based MFA, validates the OTP sent to the user's email.
- **Validate Google OTP with the Google-generated Secret**: If the user opts for Google authenticator-based MFA, validates the OTP entered by the user against the Google-generated secret key. Validation should also process on AUTH SVC.
- **Generate Backup Codes if the Request is to Enable**: If the user is enabling MFA, generates backup codes for future authentication.
- **Send Success Response**: Upon successful validation and processing of the MFA request, sends a success response to the user.

### RBAC Manage Groups Implementation LLD

CHEZ PL CIAM-PAM architecture provides mechanism to build a multi-level grouping system where individual users, for example, can be part of a company and each user needs to associated with a valid company. Then each user can be mapped to N number of companies and a company can have N number of users. Furthermore, a multi-mapping approach is inherent in the design to allow both B2B and B2B2C type of operations for an enterprise architect.

### Add/Edit Group
- **Validate Input (Condition)**: Checks if the input provided for adding/editing a group is valid.
  - inputs: name, master_id
- **Throw Validation Error (Process)**: Throws an error if the input is invalid.
- **Authorization Check (Condition)**: Verifies if the user is authorized to perform the action.
- **Throw Authorization Error (Process)**: Throws an error if the user is not authorized.
- **Add/Update Group**: Adds or updates the group with the provided master ID.
- **Send Success Response (Process)**: Sends a success response after successfully adding/editing the group.

### Delete Group
- **Validate Input (Condition)**: Checks if the input provided for deleting a group is valid.
  - group_id, master_id
- **Throw Validation Error (Process)**: Throws an error if the input is invalid.
- **Authorization Check (Condition)**: Verifies if the user is authorized to perform the action.

- **Throw Authorization Error (Process):** Throws an error if the user is not authorized.
- **Check Group Members (Condition):** Checks if the group has any members.
- **Throw Members Present Error (Process):** Throws an error if group members are present.
- **Check Permissions (Condition):** Checks if the group has any permissions assigned.
- **Throw Permissions Present Error (Process):** Throws an error if permissions are present.
- **Delete Group:** Deletes the group from the database.
- **Send Success Response (Process):** Sends a success response after successfully deleting the group.

**Group Members & Permissions - Add/Delete Group Members**
- **Validate Input (Condition):** Checks if the input provided for adding/deleting group members is valid.
    - memberId, groupId and master_id
- **Throw Validation Error (Process):** Throws an error if the input is invalid.
- **Authorization Check (Condition):** Verifies if the user is authorized to perform the action.
- **Throw Authorization Error (Process):** Throws an error if the user is not authorized.
- **Add/Delete Group Members:** Adds or deletes group members with the provided group ID and user ID.
- **Send Success Response (Process):** Sends a success response after successfully adding/deleting group members.

**Add Group Permissions**
- **Validate Input:** Checks the validity of the input parameters provided for adding group permissions.
    - group_id, permission_id
- **Throw Validation Error:** If the input parameters are invalid, an error is thrown, and the process stops.
- **Authorization:** Determines whether the user requesting to add group permissions is authorized to perform this action and, if the user type is "user", checks if the given permission ID is supposed to be added for the user role; if not, throws an authorization error and stops the *process*.
- **Throw Authorization Error:** If the user is not authorized, an error is thrown, and the process stops.
- **Add Group Permissions:** If the validation and authorization are successful, the group permissions are added to the system.
- **Send Response:** Upon successful addition of group permissions, a success response is sent to the user.
- **Stop:** Marks the end of the process.

And, additionally, the permissions table should be externally provided. For instance, for the client creation action, the permission module should be "client", the action should be "create", and the apps field can contain an array of app names. Furthermore, this permission can only be assigned to users by users if the permission role is set to "user".

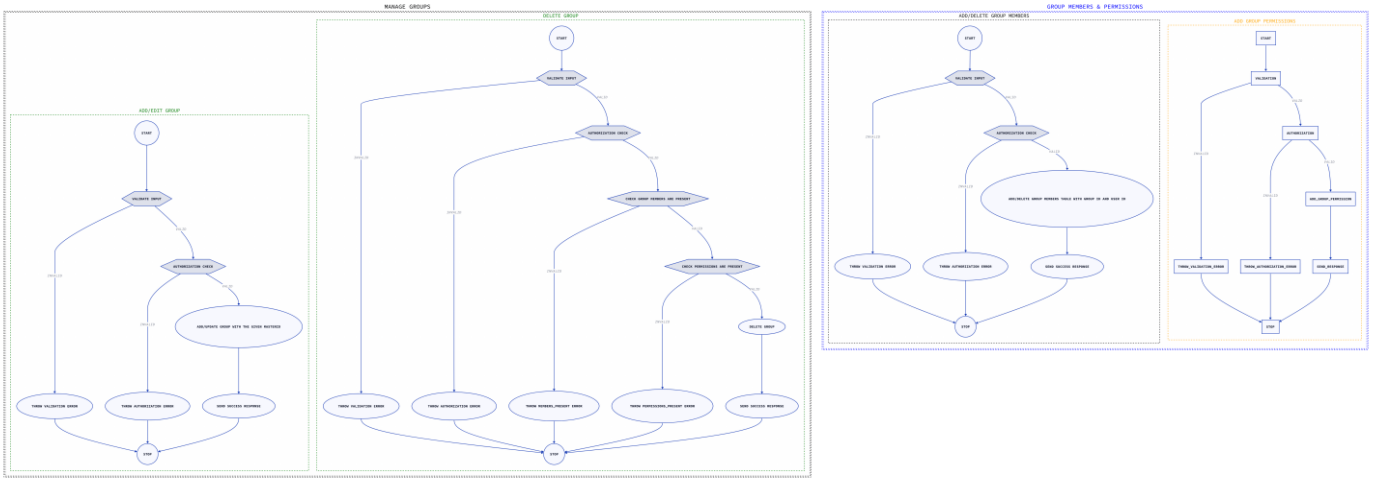

*Database Modelling Implementation*
The data model for RBAC within a CIAM system is designed to ensure efficient access control by optimizing lookups of users, their assigned roles, and associated permissions. It also focuses on maintaining data integrity through well-defined relationships and constraints, while supporting scalability to manage a large number of users, roles, permissions, and resources. Furthermore, the model facilitates auditing through proper logging mechanisms, enabling tracking of user access and permission changes and includes the ability to be future proofed, by allowing for new entities such as organizational units, to be added, ensuring it supports the changing requirements of an organization.

- **master:**
    - Serves as the main entry point or starting point in the system.
    - Indexing Strategy: Primary key index on id column.
    - Unique Constraints:
        - id: Primary key, unique key.
- **user:**
    - Represents users of the system.
    - Indexing Strategy: Primary key index on id column.
    - Unique Constraints:
        - id: Primary key, unique key.
    - Relationships:
        - Each user belongs to one master (foreign key: master_id).
- **user_details:**
    - Stores additional details about users.
    - Contains sensitive information such as passwords and verification status.
    - Indexing Strategy: Primary key index on id column. Index user_id, email, and phone columns for efficient user lookup.
    - Unique Constraints:
        - id: Primary key, unique key.
        - user_id: Unique key.
    - Relationships:
        - Each user_details belongs to one user (foreign key: user_id).
- **address:**
    - Stores user addresses.
    - Indexing Strategy: Primary key index on id column. Index user_id for efficient address retrieval per user.
    - Unique Constraints:
        - id: Primary key, unique key.
    - Relationships:
        - Each address belongs to

- one user (foreign key: user_id).
- Each user can have multiple addresses user (foreign key: user_id).
- Unique Constraints:
  - (module + action): Ensures that each combination of module and action is unique, preventing duplicate permissions.

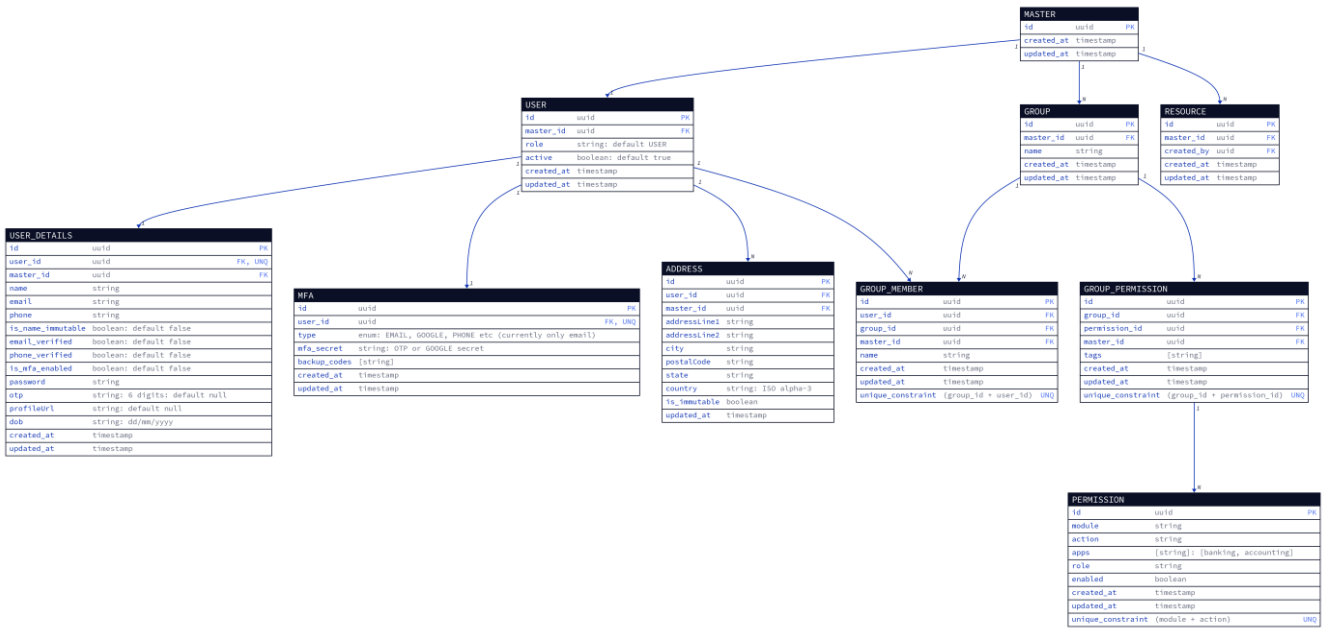

- **group**:
  - Represents groups in the system.
  - Indexing Strategy: Primary key index on id column.
  - Unique Constraints:
    - id: Primary key, unique key.
  - Relationships:
    - Each group belongs to one master (foreign key: master_id).
- **group_members**:
  - Description: Represents the membership of users in groups.
  - Indexing Strategy: Composite primary key index on (group_id, user_id) columns for efficient membership lookup.
  - Unique Constraints:
    - (group_id, user_id): Composite unique key to ensure each user is a member of a group only once.
  - Relationships:
    - Each group member belongs to one group (foreign key: group_id).
    - Each group member is a user (foreign key: user_id).
- **permission**:
  - Represents permissions granted to roles for performing actions on specific modules within the system.
  - Indexing Strategy: Primary key index on id, module column.
  - Attributes:
    - module: Indicates the module to which the permission applies (e.g., "accounting_client", "accounting_expense").
    - action: Specifies the action permitted by the permission (e.g., "list", "view", "create", "update", "delete").
    - apps: Specifies the applications associated with the permission (e.g., "banking", "accounting").
    - role: Defines the role to which the permission is granted (e.g., "admin", "user").
    - enabled: Indicates whether the permission is currently enabled or disabled.
- **group_permission**:
  - Represents the mapping between groups and permissions, defining which permissions are assigned to each group.
  - Indexing Strategy: Primary key index on id, group_id column.
  - Attributes:
    - tags: Indicates the tags associated with the group permission. This attribute is used in Attribute-Based Access Control (ABAC) systems to define additional attributes or metadata for the group permission.
  - Unique Constraints:
    - (group_id + permission_id): Ensures that each combination of group and permission is unique, preventing duplicate assignments.
  - Relationships:
    - Each group permission belongs to one group (foreign key: group_id).
    - Each group permission corresponds to one permission (foreign key: permission_id).
    - Each group permission is associated with one master record (foreign key: master_id).
- **resource**:
  - Represents resources in the system.
  - Indexing Strategy: Primary key index on id column. Index master_id for efficient resource lookup per entity.
  - Unique Constraints:
    - id: Primary key, unique key.
  - Relationships:
    - N/A

## 2. Gateway & PEP

**Gateway** acts as the central entry point for all customer requests, providing a unified interface and delegating authentication to the CIAM system. The main functions are to: route the request to appropriate backend services, validate user authentication tokens, manage microservices version & plug-ins for different applications enterprise & customer apps) via Auth02.0. Additionally, the Gateway tracks requests, responses, and traffic logs to improve observability.

On the other hand, **Policy Enforcement Point (PEP)**, manages access control decisions, based on pre-defined authorization policies. It evaluates requests, using attribute-based access control (ABAC), considering user, resource, and environmental attributes. The PEP centralizes policy enforcement, facilitating externalized authorization and supports comprehensive auditing capabilities, logging every access decision. The PEP's ability to be deployed as a separate service or to be integrated within the gateway or an application itself, highlights its versatility. **Gateway and PEP** services are designed to function together, where customer applications send requests to the Gateway, which authenticates the user's token using the Auth management system. **Gateway** forwards the request to the corresponding backend service, and then to the **PEP** service which in-turn determines if the user is authorized based on established policies. Once authorized, the Gateway allows the user access and forwards the response. The synergy between the Gateway and PEP ensures a secure, controlled, and manageable access flow within the CIAM architecture, centralizing access management decisions.

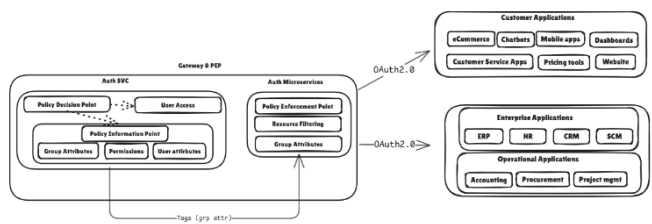

*Gateway & PEP Implementation LLD*
In CHEZ PL CIAM-PAM, a microservices architecture is followed so that challenges of scalability and stability can be achieved in a large enterprise implementation.
*Microservices:*
- *PEP (Policy Enforcement Point)*: Represents the component within microservices responsible for enforcing access control policies.
- *fetch_resource (Filter the resource with group attributes)*: Represents the action of fetching a resource and filtering it based on group attributes.

*Auth SVC (Authentication Service):*
- *PDP (Policy Decision Point)*: Represents the component responsible for making access control decisions based on defined policies.
- *PIP (Policy Information Point)*: Provides policy-related information to the Policy Decision Point (PDP).
- *Permissions*: Stores information about permissions granted to users or groups.
- *User Attributes*: Stores attributes related to users, such as master id, user role. etc
- *Group Attributes*: Stores attributes related to groups, such as group memberships or permissions.
- *Can User Access*: Represents the decision point where the authentication service determines whether a user can access a resource based on policies and attributes.

*Flow of Actions:*
1. *Microservices.PEP -> Auth SVC.PDP (1)*: The Policy Enforcement Point within microservices forwards access control requests to the Policy Decision Point in the authentication service.
2. *Auth SVC.PDP -> Auth SVC.PIP (2)*: The Policy Decision Point consults the Policy Information Point to gather necessary policy-related information.
3. *Auth SVC.PIP -> Auth SVC.Permissions, Auth SVC.Group Attributes, Auth SVC.User Attributes*: The Policy Information Point retrieves information about permissions, group attributes, and user attributes required for access control decisions.
4. *Auth SVC.PDP -> Auth SVC.Can User Access (3)*: Based on the gathered information, the Policy Decision Point determines whether the user can access the requested resource.
5. *Auth SVC.Can User Access -> Microservices.PEP (YES/NO with Group Attributes (Tags)) (4)*: The authentication service communicates the access decision (YES/NO) along with relevant group attributes back to the Policy Enforcement Point within microservices.

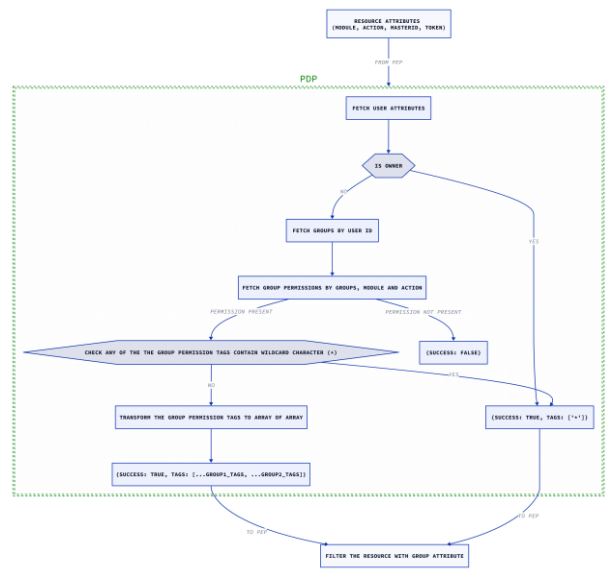

6. *Microservices.PEP -> Microservices.Fetch Resource (5)*: If access is granted, the microservices fetch the requested resource and filter it based on group attributes before providing it to the user.

*Group Attributes:*
*Tags: The user can be part of multiple groups with their own tags. For instance, if the user belongs to Group 1 with tags "Marketing1" and "Marketing2", and Group 2 with tag "Marketing3", the combined tags would be [marketing1, marketing2, marketing3]. In this scenario, the user can access resources tagged as "Marketing1", "Marketing2", or "Marketing3", effectively representing an OR condition for access permissions.*

### 3. Auth Management & PAM
Authentication Management core components include, **PIP (Policy Information Point)**: gathers user attributes and contextual information for access decisions; **PDP (Policy Decision Point)**: evaluates policies and makes access decisions; **PAP (Policy Administration Point)**: manages and defines access control policies; and **Policy & Entitlement Store**: central repository for identity policies, permissions, and trust levels.

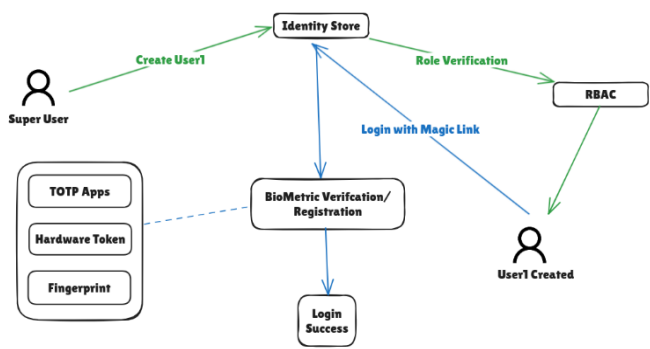

The authentication and authorization process begins with Identity and User Creation, where a Super User defines User1 by creating their account within the Identity Store, and assigns specific roles and permissions. Upon creation, a Role Verification step ensures the user's role aligns with the predefined Role-Based Access Control (**RBAC**) system. The login process utilizes Authentication with mechanisms such as **Time-Based One-Time Password (TOTP)** apps, hardware tokens, or biometric methods. Once successfully authenticated, User1 is granted access to the system.

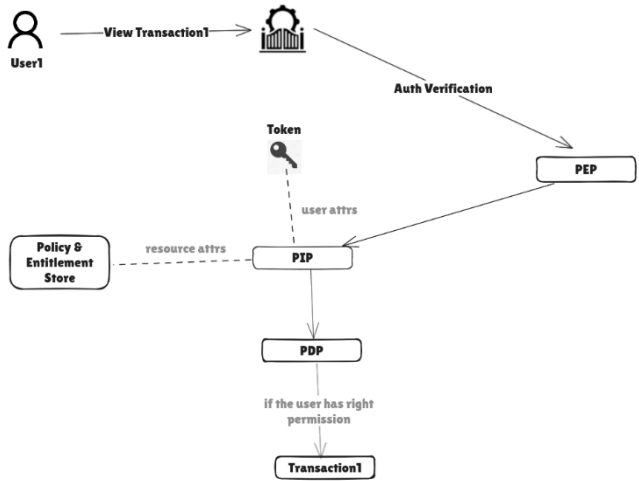

The authorization process is managed through a series of interconnected components. The **Policy Enforcement Point (PEP)** is responsible for validating user authentication. The **Policy Information Point (PIP)** then supplies the necessary user and resource attributes for decision-making. Subsequently, the **Policy Decision Point (PDP)** determines whether User1 has the required permission to access a specific resource, such as Transaction1. This entire process is underpinned by a **Policy & Entitlement Store**, where all policies and permissions associated with resources are stored.

*Authentication Management LLD Description*
*This description focuses on the authorization flow when a user attempts to access a protected resource. It assumes that authentication has already been successfully completed.*
*Request Interception by the Policy Enforcement Point (PEP):*
- **Action:** *The process begins when the PEP intercepts a request for a protected resource from an authenticated user. This request typically comes from an API Gateway, application server, or a microservice.*
- **Data:** *The PEP receives the following:*
    o *User identifier or token (obtained after authentication).*
    o *Resource identifier (e.g., API endpoint, application URL).*
    o *Requested action (e.g., read, write, delete).*
    o *Any relevant context or environment data, if present.*

*User and Resource Attribute Retrieval by the Policy Information Point (PIP):*
- **Action:** *The PEP sends a request to the PIP to gather necessary attributes for making an authorization decision.*
- **Data Flow:**
    o *The PIP receives the user identifier or token from the PEP.*
    o *The PIP queries relevant data sources (e.g., user directories, databases, external services) to retrieve the following attributes:*
        - **User Attributes:** *Role, group memberships, department, location*
        - **Resource Attributes:** *Type, owner, sensitivity, classification, etc.*
        - **Contextual Attributes**: *Time, location etc.*
    o *The PIP returns these attributes to the PEP.*

*Policy Retrieval from the Policy & Entitlement Store:*
- **Action:** *The PEP requests relevant policies associated with the resource from the Policy & Entitlement Store.*
- **Data Flow:**
    o *The PEP uses the resource identifier to query the Policy & Entitlement Store.*
    o *The Policy & Entitlement Store fetches policies that match the resource identifier and any relevant user roles.*
    o *The Store returns the applicable policies to the PEP. Policies are typically represented as structured data (e.g., JSON, XML) and define rules for granting or denying access based on attributes.*

*Authorization Decision by the Policy Decision Point (PDP):*
- **Action:** *The PEP forwards the user and resource attributes along with the policies to the PDP to evaluate the access request.*
- **Data Flow:**
    o *The PDP receives the following data:*
        - *User attributes (from PIP).*
        - *Resource attributes (from PIP).*
        - *Applicable policies (from the Policy & Entitlement Store).*
    o *The PDP evaluates the policies based on the provided attributes. This evaluation may involve complex logic, attribute comparisons, and policy precedence rules.*
    o *The PDP makes an authorization decision: either "permit" (allow access) or "deny" (block access), along with a reason code.*
    o *The PDP returns the authorization decision along with any optional context to the PEP.*

*Policy Enforcement by the PEP:*
- **Action:** *The PEP receives the authorization decision from the PDP and acts accordingly.*
- **Data Flow:**
    o *If the decision is "permit", the PEP allows the user to access the resource.*
    o *If the decision is "deny", the PEP blocks access to the resource, sending an error code back to the client.*
    o *The PEP optionally logs the authorization attempt and the access decision for auditing purposes.*

*Response to the Client:*
- **Action:** *The PEP (or the service it's a part of) returns the final response to the requesting client, based on the access decision.*
- **Data:**
    o *If access is granted, the client receives a successful response and access to the protected resource.*
    o *If access is denied, the client receives an access denied message or error.*

*Key Considerations in the LLD:*
- **Data Format and Protocols:** *Define the data formats for communicating between the PEP, PIP, and PDP (e.g., JSON, XML) and the protocols used for communication (e.g., REST, gRPC).*
- **Caching:** *Implement caching mechanisms in the PIP to reduce the overhead of retrieving attributes repeatedly and also for policies, to reduce the overhead from the store.*
- **Policy Language:** *Choose a policy language (e.g., XACML, OPA) that can handle the complexity of your requirements and provides the flexibility for policy expression.*

- **Error Handling:** *Define how errors will be handled at each stage and how they are communicated back to the client.*
- **Logging and Auditing:** *Implement logging mechanisms to track authorization requests, decisions, and any policy evaluation errors.*
- **Scalability and Performance:** *Consider scalability and performance when designing the PEP, PIP, PDP, and Policy & Entitlement Store, ensuring they can handle the expected load and performance requirements.*
- **Security:** *Ensure that all internal communication is secure and that there is protection for the policies in the store, and that data is handled according to compliance requirements.*

**Privileged Access Management (PAM)**

The Privileged Access Management (PAM) system comprises several key components. At its core is the *Vault*, a centralized repository for storing sensitive credentials such as SSH keys, LDAP credentials, and access keys. The *Privileged Session Manager (PSM)* monitors privileged sessions, ensuring security and compliance. The *Infrastructure*, consisting of Servers and Databases, is secured by the PAM system. To mitigate the risks of unauthorized access, the system utilizes *Password Rotation*, an automated process for the frequent update of passwords. *Session Monitoring* further enhances security through the tracking and recording of user sessions, with logs being sent to external systems like Splunk/Audit Logs for analysis and compliance reporting.

In a combined CIAM and PAM architecture, the **Vault** is more than just a password repository. It's a central control point for managing all types of sensitive credentials across diverse user types and system environments. It provides a secure, auditable, and manageable solution for credential management, that enhances overall security by centralizing storage, automating rotation, and enforcing access control policies. The Vault's tight integration with CIAM and PAM systems enables seamless access management and provides a unified platform for credential management. This integration simplifies operations, reduces the risk of unauthorized access, and strengthens the overall security posture of the combined CIAM/PAM architecture.

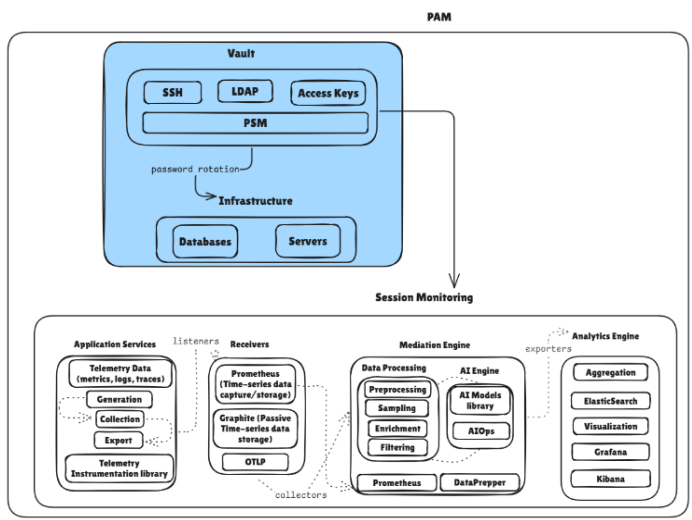

**Centralized Credential Storage:** The Vault provides a single, secure location for storing all types of sensitive credentials required by both CIAM and PAM systems. This includes:

- **PAM Credentials:**
    - **SSH Keys:** For accessing servers, network devices, and other infrastructure components. These keys are used for passwordless authentication for privileged users and accounts.
    - **LDAP Credentials:** For accessing directory services, enabling authentication and authorization for both users and applications. These may include service accounts.
    - **Access Keys:** For accessing cloud environments, databases, and APIs, providing programmatic access to different resources. These keys are often used by system administrators and developers.
    - **Privileged Account Passwords:** For accessing critical systems, databases, and applications that require elevated privileges.
- **CIAM Credentials:**
    - **API Keys:** For applications to access customer-facing APIs, allowing for secure interaction between applications and the core CIAM system.
    - **Secret Keys:** Used for OAuth 2.0 clients, and other secure communication.
    - **Service Account Passwords:** For backend processes or services, allowing secure access to resources without the need for human intervention.
- **Credential Management:** Beyond just storage, the Vault manages the lifecycle of these credentials:
    - **Secure Creation and Storage:** Credentials are created and stored using strong encryption and secure access control mechanisms.
    - **Automated Rotation:** The Vault automates the rotation of passwords and other credentials, reducing the risk of compromise due to credential leaks or reuse. This ensures that credentials are changed frequently, minimizing the impact of any security breach.
    - **Access Control:** Access to the Vault itself is tightly controlled, typically through role-based access control (RBAC) and/or attribute-based access control (ABAC) policies, ensuring only authorized users and systems can retrieve sensitive credentials.
    - **Policy Enforcement:** The Vault enforces defined policies regarding the usage of credentials such as access times, IP address restrictions, and session duration.

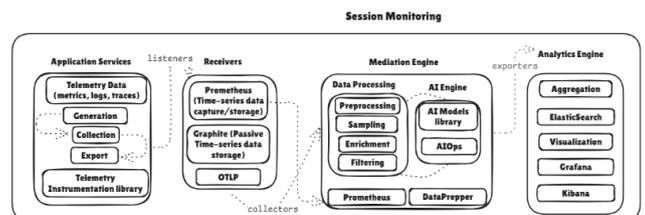

- **Integration with CIAM and PAM Workflows:**
    - **PAM Integration:** The Vault seamlessly integrates with the PAM system, providing the required credentials to the *Privileged Session Manager (PSM)* for privileged user access. The Vault provides the necessary credentials for user authentication and session establishment, and ensures that users authenticate with a multi-factor authentication (MFA) method before accessing the vault.
    - **CIAM Integration:** The Vault securely provides API keys and service account credentials to applications and services within the CIAM ecosystem. This provides a way to secure access between internal systems.
- **Enhanced Security Posture:**
    - **Reduced Attack Surface:** By centralizing credentials, the attack surface is reduced, simplifying security management and preventing credentials from being embedded in application

- code or configuration files.
  - **Compliance Support:** The Vault supports compliance with industry standards and regulations (e.g., PCI DSS, GDPR) by maintaining a complete audit trail of credential access and changes.
  - **Auditing:** The Vault tracks all the activity on how each credential is used and can generate logs for audit reports.
- **Logical Separation of Customer and Privileged Credentials:**
  - While the Vault provides centralized storage, it can also logically separate storage for customer-facing CIAM credentials and internal-facing privileged PAM credentials. This separation ensures an additional layer of security, limiting access and segregation of duty.
  - The vault also logically separates storage of credentials based on environment, separating production, testing and development credentials for example.

**Automated password rotation** is a core security mechanism, designed to eliminate manual password changes through scheduled and event-triggered updates. This process uses randomized password generation to ensure strong and unique credentials, applied across a broad spectrum of privileged targets, including server, database, application, service, and cloud account passwords. The system integrates with the **PAM Vault** for secure storage and controlled credential distribution, while leveraging the **PSM** to ensure seamless user experience. This automated approach enhances the security posture through a reduction in the attack surface, minimized lateral movement, and improved compliance. Additionally, it provides operational efficiency with reduced manual effort and a consistent process for password changes.

**End-to-end Data Encryption**
Data transmitted with TLS/SSL is protected during transit but requires server-side decryption for processing and routing, creating an attack vector. Likewise, while encryption at rest safeguards physical storage, compromised accounts with decryption access can bypass it. This results in plaintext exposure, even momentarily, on intermediary systems like load balancers, proxies, and API gateways, forming another attack vector. An attacker exploiting these vulnerabilities can compromise sensitive IAM data, such as user credentials and access tokens. Furthermore, malicious insiders within the IAM infrastructure have the capability to decrypt and misuse this sensitive information. These vulnerabilities highlight the imperative for a more resilient approach that safeguards data, even if an intermediary system is breached.
- **Client-Side Encryption:**
  - Encryption of identity and authentication data is performed at the client's device or application.
  - Ensures that keys are not exposed to intermediaries.
- **Secure Key Management:**
  - Secure and effective management of cryptographic keys is critical.
  - User and session-specific key derivation reduces the impact of key compromise.
  - Robust key exchange mechanisms, such as Diffie-Hellman, are essential.
- **Encrypted Authentication/Authorization:**
  - E2EE protects all authentication and authorization requests in transit.
  - Critical identifying information remains encrypted even if intermediaries are compromised.
- **Secure Data Exchange with IAM Service:**
  - Only the intended IAM service decrypts and processes requests.
  - All responses from the IAM service are encrypted before transmission.
  - The IAM service must support and utilize E2EE.
- **Minimal Trust in Intermediaries:**
  - E2EE drastically minimizes the necessary trust placed in intermediary systems.
  - Intermediaries simply transport encrypted data without access to plaintext.

**4. Session Monitoring**
The Session Monitor system provides real-time traffic capture, correlation, and indexed storage for subsequent reporting through a web interface. The system architecture comprises three distinct layers: a *Probe Layer*, responsible for capturing network traffic and performing media quality analysis, sending signaling metadata and RTP analysis results to the Mediation Engine; a *Mediation Engine Layer*, which processes, correlates, and stores the received traffic for future analysis, and manages key performance indicators (KPIs), typically deployed on a geographical site basis; and an *Aggregation Engine Layer*, responsible for aggregating global KPIs from multiple Mediation Engines and supporting global search functionalities, which is usually a single instance for the entire network.

This architecture is designed for comprehensive, intelligent monitoring of session data, leveraging AI for enhanced insights and proactive issue detection. Here's a breakdown of each component:

**Application Services:**
- **Role:** Application services represent the various applications and systems within your environment that generate session data. These could be web applications, mobile apps, microservices, databases, or any other system where user activity occurs.
- **Telemetry Data (Generation, Collection, Export):**
  - **Generation:** The application services generate telemetry data, which includes metrics, logs, and traces related to user sessions. This data could be performance metrics (e.g., response times, latency), security events (e.g., login attempts, permission violations), usage patterns, and other relevant data.
  - **Collection:** Telemetry data is collected using a variety of methods. This might include instrumenting the application code directly, using agent-based collection, or leveraging log aggregators.
  - **Export:** Collected telemetry data is exported from the application services to receivers, typically using protocols like HTTP, gRPC, or TCP.
- **Telemetry Instrumentation Library:**
  - **Purpose:** This library is embedded within the application services to facilitate the generation, collection, and export of telemetry data.
  - **Functionality:** The library provides APIs for emitting metrics, traces, and logs, and handles the formatting and transmission of this data to configured destinations. It offers features like automatic context propagation, sampling, and batching to optimize data collection efficiency.
- **Listeners:**
  - **Purpose:** The application services use listeners to establish connections with receivers to transmit the collected telemetry data.
  - **Functionality:** The listeners are configured to listen on specified ports or addresses where the receivers are listening for data.

**Receivers:**
- **Role:** Receivers act as intermediaries that receive telemetry data exported from the application services and prepare it for further processing. They handle different data formats and protocols, and provide standardized access to the telemetry data.
- **Types:**
    - **Prometheus:** A widely used open-source monitoring and alerting system that primarily collects active time-series data and metrics. Receives metrics data and can process it to be used by the mediation engine.
    - **Graphite:** A open-source monitoring system focused on storing passive time-series data.
    - **OTLP (OpenTelemetry Protocol):** A standard protocol for telemetry data, that supports metrics, logs and traces and allows

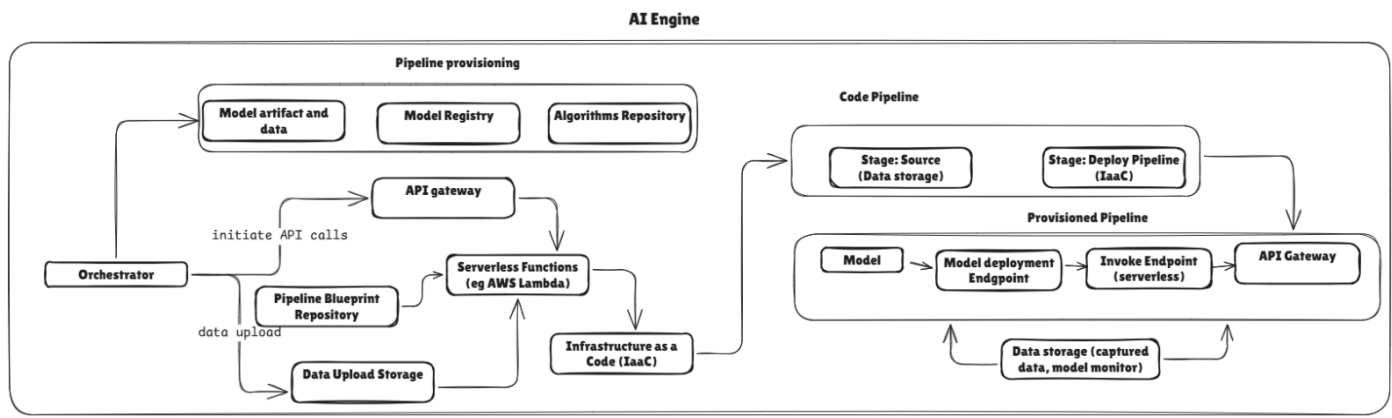

for standardized collection and transmission to the mediation engine.
- **Collectors:**
    - **Purpose:** Collectors are components that collect data from the receivers, and then transmit it to the mediation engine.
    - **Functionality**: The collectors are configured to listen on specified addresses and ports where the receivers are transmitting data.

**Mediation Engine (Data Processing & AI Engine):**
- **Role:** The Mediation Engine is the core processing component responsible for data ingestion, transformation, and enrichment, as well as providing intelligent insights through AI-driven analysis.
- **Data Processing:**
    - **Data Ingestion:** Receives data streams from multiple receivers via collectors.
    - **Data Transformation:** Transforms data into a consistent format, standardizing metrics, logs, and traces for subsequent analysis.
    - **Data Enrichment:** Enriches data with additional context, such as user attributes, location information, or other relevant data obtained from external systems.
- **AI Engine:**
    - **Anomaly Detection:** Employs machine learning models to detect unusual patterns or behaviors in session data, such as sudden spikes in latency or an increase in failed login attempts.
    - **Root Cause Analysis:** Analyzes collected data to identify the underlying causes of performance bottlenecks or security incidents.
    - **Predictive Analytics:** Uses historical data to forecast potential issues, such as capacity limits or future security threats.

- **Prometheus & Dataprepper**
    - **Purpose**: Prometheus collects metrics from the mediation engine itself, while Data Prepper acts as a buffer, filtering and transforming data for processing.
    - **Functionality**: Prometheus is used to monitor the mediation engine itself, while DataPrepper is an optional processing engine that can provide additional processing for the mediation engine's data, and also buffer any issues with the connection to the analytics engine.

**4. Analytics Engine:**
- **Role:** The Analytics Engine stores and analyzes processed data from the mediation engine, providing a platform for data visualization and reporting.
- **Components:**
    - **Aggregation:** Aggregates the data received from the mediation engine, often into time-based buckets or categories.
    - **Elasticsearch:** A distributed, open-source search and analytics engine used for indexing and querying large volumes of data, which allows for performant searches across large amounts of log data.
    - **Grafana/Kibana:** Visualization dashboards (Grafana & Kibana) provide interactive views of the aggregated data, with customizable widgets for monitoring key performance indicators (KPIs) and identifying potential issues. Provides a method for monitoring the overall state of the system.

**Data Flow for Session monitoring:**
- **Data Generation:** Application services generate telemetry data.
- **Data Export:** Telemetry data is exported through a library, via listeners, to receivers (Prometheus, Graphite, OTLP).
- **Data Collection:** Collectors receive and forward telemetry data to the mediation engine.
- **Data Processing & AI Analysis:** The mediation engine processes the telemetry data, applying AI models for analysis.
- **Processed Data Storage and Monitoring**: The results are forwarded to Prometheus and DataPrepper for buffer, filtering and transformation for the analytics engine.
- **Data Aggregation:** The Analytics Engine aggregates data and indexes it in elasticsearch.
- **Data Visualization:** Grafana/Kibana visualize data from elasticsearch.

**AI-Powered Analytics and Monitoring**
This AI-powered session monitoring architecture provides a holistic approach to understanding system behavior. Application services capture telemetry data, which is then processed and analyzed by the Mediation Engine, and then aggregated, stored and visualized using the Analytics Engine. The system utilizes AI capabilities to provide intelligent insights, proactive issue detection, and detailed reporting, and is designed to be scalable, reliable, and adaptable to the evolving needs of modern systems.

## V. FUTURE EXTENSIONS

The integration of AI and GenAI into Zero Trust CIAM architectures holds immense potential for enhancing security, improving user experiences of security processes, automating security administrative tasks, and automated support for data privacy. This research section is intended to explore those opportunities, while also recognizing and addressing the ethical and technical challenges that must be overcome to fully realize these benefits. Further research is essential to develop practical, robust, and responsible AI and GenAI solutions that can meet the evolving needs of modern digital ecosystems.

**Adaptive Authentication with AI:**
- **Current Limitations:** Traditional multi-factor authentication (MFA) can be rigid and user-intrusive.
- **AI Solution:** AI can analyze various factors in real-time (user behavior, device posture, network location) to dynamically adjust authentication requirements. For low-risk actions, users might experience seamless access, while higher-risk actions trigger stricter verification, reducing user friction while maintaining security.
- **Example Use Cases:** Seamless login based on behavioral biometrics, adaptive MFA based on risk score.

**AI-Driven Anomaly Detection:**
- **Current Limitations:** Rule-based anomaly detection can miss sophisticated threats.
- **AI Solution:** AI can learn normal user behavior and identify unusual patterns in login attempts, access patterns, or data access, providing early alerts for suspicious activities.
- **Example Use Cases:** Identification of compromised accounts based on anomalous login patterns, early detection of data exfiltration attempts.

**Automated Identity Governance with AI:**
- **Current Limitations:** Manual identity governance processes are time-consuming and prone to errors.
- **AI Solution:** AI can automate tasks such as user provisioning, de-provisioning, and role assignment based on user attributes and organizational structure. AI can continuously analyze user activity to identify role discrepancies and suggest policy adjustments.
- **Example Use Cases:** Automatic role adjustment for users based on their job function, risk-based access reviews.

## VI. LIMITATIONS

This section outlines the limitations associated with an AI-driven Zero Trust CIAM-PAM combined architecture, highlighting potential implementation hurdles, technical constraints, and ethical concerns. Acknowledging these limitations is crucial for future research and development efforts to build robust and reliable systems. Further investigation and mitigation strategies are needed to address these limitations, ensuring that the benefits of AI-driven security can be fully realized in real-world environments. Future research should focus on creating solutions for these issues, allowing us to fully harness the potential of the AI-driven zero-trust architecture.

**Complexity and Implementation Challenges:**
- **Integration Complexity:** Integrating diverse CIAM and PAM systems, especially with the addition of AI components, can be highly complex. Ensuring seamless interoperability, data consistency, and policy enforcement across all layers presents a significant challenge.
- **Data Silos:** Combining data from customer-facing and privileged user environments can be challenging due to differing data formats, schemas, and access control policies, creating silos that impede effective AI analysis.
- **Deployment Complexity:** The complex nature of AI algorithms, and the need to customize them for different scenarios, makes deployment complicated. Setting up and fine-tuning AI components can be resource-intensive and require specialized expertise.
- **Resource Intensive:** AI and GenAI models, especially with real-time analysis requirements, can be resource-intensive, requiring significant compute, memory, and storage resources, potentially incurring high operational costs.

**Reliance on AI and Potential Biases:**
- **Data Bias:** AI models are trained on data, and if the training data is biased, the AI system can perpetuate and even amplify these biases, leading to discriminatory access control decisions, and potentially impacting user experiences.
- **Model Opacity:** Many AI models operate as "black boxes," making it difficult to understand their decision-making processes. This lack of transparency can hinder auditability and trust in the system, especially for high impact access decisions.
- **Adversarial Attacks:** AI models can be vulnerable to adversarial attacks designed to deceive or manipulate them, potentially bypassing security measures. These attacks can be difficult to detect and mitigate.
- **False Positives and Negatives:** The system can generate false positives, triggering unnecessary alerts, or false negatives, where real threats go undetected.
- **Over-reliance on Automation:** Over-reliance on AI-driven automation may cause a reduced ability to respond to novel security challenges.

**Scalability and Performance Concerns:**
- **Real-time Performance:** Processing and analyzing large volumes of session data in real-time can be challenging. Complex AI computations may introduce latency, impacting the overall system performance and user experience.
- **Scalability Limitations:** The architecture may face scalability issues as the number of users, applications, and resources increases. Scaling AI components in a distributed environment can be complex and require careful planning.
- **Maintenance of Models**: The AI/GenAI models will require continuous retraining and updates to ensure they continue to work in the changing environment.

**Data Privacy and Compliance:**
- **Data Collection and Usage:** Collecting and processing user data for AI analysis raises privacy concerns and requires compliance with data protection regulations (e.g., GDPR, CCPA). Transparency about how data is used is crucial.

- **Data Security:** Protecting sensitive data used in AI models from unauthorized access and breaches is paramount. Data encryption, anonymization, and secure storage practices are essential.
- **Policy Compliance:** The architecture must be designed to meet a variety of different regulatory compliance requirements; this can be complex to achieve in a system with changing AI models.

**Lack of Standardization and Maturity:**
- **Evolving Technologies:** The technologies used in AI-driven security are constantly evolving. The lack of clear standards and best practices can make it challenging to implement interoperable and future-proof solutions.
- **Limited Research and Testing:** This technology is still in its early stages of development, and lacks rigorous testing, and has limitations that are yet to be discovered.
- **Vendor Lock-In:** Reliance on proprietary AI solutions may lead to vendor lock-in, limiting flexibility and potentially increasing costs.

**Human Factor and Operational Challenges:**
- **Skill Gap:** Implementing and managing such systems requires specialized expertise in AI, cybersecurity, and identity management. A shortage of skilled personnel can create implementation challenges and affect long-term viability.
- **Adaptation to New Processes:** Organizations must adapt to new processes and workflows introduced by AI-driven systems. User training and change management are critical.
- **Maintenance Overhead:** Continuous monitoring and maintenance of the AI components requires skilled personnel and robust maintenance policies.

## VII. RESULTS

This research demonstrates the efficacy of the proposed CHEZ PL CIAM-PAM architecture in addressing the limitations of traditional IAM systems within the context of evolving cybersecurity threats. Key results include the successful implementation of adaptive authentication, identity federation, and AI-powered continuous session risk analysis, demonstrating enhanced security and scalability compared to legacy approaches. The architecture's Zero Trust framework enables granular access control and adaptive policy enforcement, overcoming the constraints of static, RBAC systems. Further, the incorporation of enhanced MFA, Authentication management, federated identity management, integrated privileged user management and advanced behavioral analytics supports distributed environments and multi-platform integration while adhering to compliance standards such as GDPR, HIPAA, and SOC 2. Notably, the microservices-based design of the CHEZ PL model ensures seamless interoperability with legacy systems. Finally, this research successfully incorporates AI-driven session monitoring and offers automated reporting for compliance. This new approach is shown to be adaptable to emerging AI and GenAI technologies, thereby presenting a future-proof solution for large multi-national organizations.

- **Zero-Trust Principles:** The CHEZ framework adheres to the zero-trust model by treating every access request as untrusted until verified. This approach eliminates implicit trust and minimizes attack surfaces.
- **Micro-Segmentation:** The architecture implements micro-segmentation to isolate sensitive systems and reduce lateral movement in case of a breach.
- **Data Encryption Standards:** All data, including identity attributes and privileged credentials, are encrypted both at rest and in transit.
- **AI-Based Monitoring:** Continuous monitoring powered by AI ensures proactive threat detection and rapid response to incidents.

**Scalability and Flexibility**
- The CHEZ CIAM-PAM model is built for scalability, supporting distributed networks and hybrid environments.
- It integrates with legacy systems through API connectors, reducing the complexity of migration. Its modular design allows enterprises to customize and extend functionalities as their requirements evolve.

**AI and GenAI Integration**
- AI and GenAI tools enhance the framework by enabling predictive analytics, behavioral analysis, and adaptive policy enforcement.
- These tools are embedded into the monitoring and incident response workflows, ensuring proactive threat detection and automated mitigation strategies.

**Integration with Cloud and Hybrid Environments**
- The CHEZ architecture supports cloud-native deployments, hybrid models, and on-premise implementations. Its cloud-agnostic approach enables seamless integration with AWS, Azure, and Google Cloud platforms.
- This detailed overview of the CHEZ CIAM-PAM architecture demonstrates its capability to address modern cybersecurity challenges while maintaining compliance, scalability, and adaptability. It serves as a comprehensive framework for enterprises looking to adopt a zero-trust model with advanced AI-driven capabilities.

## VIII. CONCLUSION

This paper detailed the CHEZ PL CIAM-PAM framework, a novel architecture incorporating zero-trust principles and designed for hyper-extensibility to address contemporary enterprise security challenges. The framework's core innovations - AI-driven analytics, federated identity management, adaptive MFA, and pseudonymized data handling - achieve significant advancements in security, compliance, and scalability.

**Migration Plan: IAM Decommissioning & CIAM Implementation**
**Phase 1: Planning and Assessment**
*Step 1.1: Project Initiation and Stakeholder Alignment*

- **Objective:** Establish project goals, scope, success criteria, and identify key stakeholders.
- **Tasks:**
  - Define clear business objectives for the CIAM migration (e.g., improved user experience, enhanced security, scalability).
  - Identify project sponsors, core team members, and other relevant stakeholders.
  - Establish a governance structure and communication plan.
  - Define the overall project timeline, budget, and resource allocation.
  - Develop a risk management plan and identify potential roadblocks.

*Step 1.2: Current IAM System Assessment*

- **Objective:** Thoroughly analyze the existing IAM system and its functionalities.
- **Tasks:**
    - Document the existing IAM architecture, including components, configurations, and integrations.
    - Identify all applications, services, and users currently managed by the existing IAM system.
    - Analyze authentication methods, authorization policies, and user lifecycle processes.
    - Assess the performance, scalability, and security limitations of the current IAM.
    - Identify data dependencies and integration points with other systems.
    - Document all custom configurations, scripts, and extensions.

*Step 1.3: CIAM Solution Evaluation and Selection*

- **Objective:** Evaluate potential CIAM solutions and select the one that best meets business and technical requirements.
- **Tasks:**
    - Define CIAM requirements based on the assessment (Step 1.2) and project objectives.
    - Research and evaluate available CIAM solutions based on features, functionality, scalability, security, vendor support, and cost.
    - Conduct proof-of-concept (POC) implementations for shortlisted solutions.
    - Select the final CIAM solution based on the POC results and evaluation criteria.

*Step 1.4: Target Architecture Design*

- **Objective:** Design the target CIAM architecture, including integrations, customizations, and security controls.
- **Tasks:**
    - Design the CIAM architecture based on the selected solution and business requirements.
    - Plan integrations with target applications, databases, and other systems.
    - Define data migration strategies (e.g., full migration, incremental migration).
    - Design authentication and authorization flows within the new CIAM system.
    - Develop a comprehensive security plan for the new CIAM environment, including access controls, encryption, and monitoring.
    - Create a detailed system design document.

**Phase 2: Migration Preparation**
*Step 2.1: CIAM Solution Deployment and Configuration*

- **Objective:** Deploy and configure the selected CIAM solution according to the design specifications.
- **Tasks:**
    - Set up the new CIAM environment, including servers, databases, and other infrastructure components.
    - Configure the core CIAM functionality, including identity storage, authentication methods, and authorization policies.
    - Implement any required customization based on business requirements.
    - Set up development, staging, and production environments for CIAM.
    - Develop automated deployment processes.

*Step 2.2: Data Migration Preparation*

- **Objective:** Prepare user data and other relevant information for migration to the new CIAM.
- **Tasks:**
    - Define data mapping rules between the old and new IAM systems.
    - Extract data from the existing IAM system.
    - Cleanse and transform data to match the new CIAM data model.
    - Develop data migration scripts and procedures.
    - Plan and execute test data migrations to validate the process.

*Step 2.3: Application Integration Development*

- **Objective:** Develop integrations between target applications and the new CIAM.
- **Tasks:**
    - Develop integration interfaces and adaptors for all applications and systems that will integrate with the CIAM.
    - Implement authentication and authorization flows for each application.
    - Conduct integration testing to ensure proper functionality and security.
    - Develop documentation for integrated applications.

*Step 2.4: User Onboarding Preparation*

- **Objective:** Prepare user communication and documentation to ensure a smooth transition.
- **Tasks:**
    - Develop user guides and tutorials for the new CIAM system.
    - Plan and execute user communications, informing users of the migration timeline, changes, and any required actions.
    - Develop a help desk procedure for supporting user inquiries.

**Phase 3: Migration Execution**
*Step 3.1: Staged Migration and Testing*

- **Objective:** Migrate users and applications in phases, starting with a pilot group and progressing to larger groups.
- **Tasks:**
    - Migrate a small group of pilot users to the new CIAM system.
    - Monitor the pilot migration, address issues, and refine the migration process.
    - Migrate subsequent groups of users and applications in stages.
    - Perform thorough functional, performance, and security testing after each migration wave.

*Step 3.2: Full Data Migration*

- **Objective:** Migrate all remaining users, data, and configurations to the new CIAM system.
- **Tasks:**
  - Execute the full data migration plan using the established procedures.
  - Monitor the migration process to identify and resolve any issues.
  - Validate data integrity after the migration is complete.
  - Conduct post-migration testing to ensure all functionality is operational.

*Step 3.3: Application Cutover*

- **Objective:** Switch all applications to use the new CIAM system for authentication and authorization.
- **Tasks:**
  - Coordinate the cutover of all integrated applications.
  - Monitor the cutover process and respond to any incidents.
  - Provide support to application owners during the cutover.

*Step 3.4: Monitoring and Validation*

- **Objective:** Continuously monitor the new CIAM environment and ensure proper functionality and performance.
- **Tasks:**
  - Implement comprehensive monitoring solutions for the CIAM infrastructure and applications.
  - Validate the authentication and authorization policies after migration.
  - Perform regular security assessments and penetration tests.
  - Ensure all logs and audit trails are functioning as expected.

**Phase 4: Decommissioning and Post-Migration Support**
*Step 4.1: Existing IAM Decommissioning*

- **Objective:** Safely decommission the old IAM system, ensuring no data or services are lost.
- **Tasks:**
  - Verify that all functionality has been successfully migrated to the new CIAM solution.
  - Backup the old IAM system and all relevant data.
  - Decommission the old IAM infrastructure.
  - Dispose of hardware and software in compliance with company policies.

*Step 4.2: Post-Migration Support and Training*

- **Objective:** Provide ongoing support, training, and maintenance for the new CIAM system.
- **Tasks:**
  - Provide user support and respond to inquiries.
  - Conduct training sessions for system administrators and users on how to use the new CIAM.
  - Establish ongoing system maintenance and monitoring schedules.
  - Document all support procedures and troubleshooting steps.
  - Gather feedback from users and identify areas for improvement.

*Step 4.3: Post-Implementation Review*

- **Objective:** Assess the success of the CIAM migration and identify any lessons learned.
- **Tasks:**
  - Conduct a post-implementation review to evaluate the project against success criteria.
  - Document lessons learned and best practices for future migrations.
  - Prepare a final report summarizing the project outcomes, costs, and benefits.

The CHEZ framework effectively addresses key vulnerabilities in traditional CIAM-PAM systems by leveraging AI to provide real-time risk assessment, anomaly detection, and predictive analytics. This framework demonstrates the viability of AI-driven security within a zero-trust CIAM-PAM paradigm, facilitating dynamic access controls, enhanced session monitoring, and adaptive policy enforcement, while maintaining interoperability with legacy systems. Future research should focus on integrating GenAI for advanced behavioral analysis, developing AI-driven automated compliance checks, and investigating blockchain for decentralized identity management. Further, research must explore deployment strategies and operational performance across different industries. The CHEZ PL CIAM-PAM architecture provides a scalable, secure, and adaptive IAM system, built on a strong foundation for defending against evolving cyber threats.

## IX. FUTURE SCOPE

The future scope of the CHEZ PL CIAM-PAM framework presents numerous opportunities for further enhancements, research, and practical implementations. The integration of AI and GenAI offers significant potential for improving predictive analytics, anomaly detection, and real-time response mechanisms. Future enhancements may focus on:

- **AI-Driven Automation and Orchestration:** Expanding the use of AI algorithms to automate identity lifecycle management, policy enforcement, and dynamic threat mitigation. AI-powered bots can handle repetitive tasks, reducing manual effort and improving operational efficiency.
- **Blockchain-Based Identity Verification:** Leveraging blockchain technology to establish decentralized identity verification mechanisms, enhancing security and privacy in identity management processes.
- **Advanced Behavioral Analytics:** Implementing behavior-based anomaly detection and continuous risk assessments to predict and counter threats before they escalate.
- **Edge Computing Integration:** Incorporating edge computing for faster authentication processes in IoT-enabled environments, ensuring seamless access control in distributed systems.
- **Cross-Platform Compatibility:** Expanding the framework to support hybrid cloud environments, multi-cloud deployments, and edge computing nodes for improved scalability and performance.
- **Compliance Management Enhancements:** Integrating AI tools for automated compliance audits and reporting to meet

evolving global regulatory standards.

- **Quantum-Resistant Cryptography:** Preparing for future advancements in quantum computing by implementing encryption methods that resist quantum attacks.
- **Dynamic Access Controls with Context-Awareness:** Enhancing access control models with dynamic policy enforcement that adapts based on contextual factors such as location, device, and behavior patterns.
- **Secure Data Collaboration Models:** Developing methods for secure data sharing and collaboration across organizations while maintaining privacy through pseudonymization and encryption.
- **AI-Enhanced Governance Frameworks:** Incorporating AI-driven insights to optimize governance, risk, and compliance (GRC) processes.

These future directions will continue to push the boundaries of CIAM and PAM systems, ensuring enterprises remain resilient against evolving cybersecurity threats.